# High Entropy Alloys Mined From Phase Diagrams

Qi Jie[1], Andrew Cheung[2], and S. Joseph Poon[1]

[1] Department of Physics, University of Virginia, Charlottesville, VA 22904-4714
[2] Department of Materials Science, University of Virginia, Charlottesville, VA 22904-4259

**Abstract**

**High entropy alloys (HEA) show promise as a new type of high-performance structural material. Their vast degrees of freedom provide for extensive opportunities to design alloys with tailored properties. However, the compositional complexities of HEAs present great challenges for alloy design. Current approaches have shown limited reliability in accounting for the compositional regions of single solid solution and composite phases. We present a phenomenological method, analyzing binary phase diagrams to predict HEA phase formation on the hypothesis that the structural stability of HEAs is encoded within. Accordingly, we introduce a small number of phase-diagram inspired parameters and employ machine learning to partition the formation region of 500+ reported HEA compositions. The model achieved a single phase HEA prediction rate >80 %. To validate our method, we demonstrated the capability of this method in predicting HEA solid solution phases with and without intermetallics in 30 randomly selected complex compositions, with a success rate of 77 %. The presented efficient search approach with high predictive capability can be exploited to complement computation-intense methods in providing a powerful platform for the design of high entropy alloys.**

## Introduction

High entropy alloys (HEAs) were first discovered in 2004[1-2]. They are also known as multi principal elements (MPE) alloys or compositionally complex alloys (CCA). HEAs can form as either single or mixed phases. HEAs have emerged as one of the most popular topics in material research[1-5]. These materials span vast compositional space, providing flexibility in alloy design[6-11]. However, the compositional complexity poses a significant challenge in the control of phase formation due to thermodynamic and kinetic constraints[12,13]. Empirical approaches that utilized atomistic and thermodynamic parameters[14-17] were first introduced to investigate the compositional regions of HEA phases, but with only limited success. Additionally, first-principles calculation[16,18-22] and Calculation of Phase Diagrams (CALPHAD)[23] methods have been employed to shed light on the atomistic and thermodynamic mechanisms of HEA formation. However, the accuracy of CALPHAD is often limited by the availability of thermal databases, and the appearance of miscibility gaps and intermetallic (IM) compounds in the binary systems[24]. Monte Carlo simulations show promising results in predicting the formation of certain IM phases and the phase structure evolution with varying temperatures[25]. Employing statistical approaches, a thermodynamics and Gaussian process statistical model[26] that utilized up to nine parameters was proposed as the basis for identifying single solid solution phases. Using a database of over 2000 multicomponent alloy compositions from a high-throughput sputter deposition experiment, another model applied a regression method to interrogate the HEA phase formation tendency[27]. Despite recent progress in understanding the formation trend of subgroups of HEAs, the constitution of HEAs still relies on trial and error, which impedes the design of these multicomponent alloys for fundamental studies and applications.

High entropy provides the driving force for a HEA system to form a single solid solution phase. A distinctive feature of good HEA forming systems is significant to moderate solid solution formation tendency among the constituent binary alloys. However, the experimental scenario is more complex. For one thing, different solid solution phases can coexist, and phase separation and IM formation occur often, potentially disrupting the formation of single-phase HEAs. Other factors such as atomic interaction and atomic-level strain as well as temperature that influence phase formation and stability must be taken into account. Except for the computation intensive studies, atomic interactions are usually not comprehensively accounted for by the prior mentioned models. On the other hand, the experimentally validated phase diagrams are encoded with the binary atomic interaction information.

Departing from current approaches, we present herein a phenomenological method as the basis of reality to predict the compositional space of HEA phases. At the outset, the advantage of using binary phase diagrams to assess phase stability is that they can readily provide direct and realistic information about the roles of individual elemental components on phase formation. Albeit only for binary alloy systems, nevertheless, the phenomenological method is built on the hypothesis that the constituent binary alloys encode a wealth of information about the multicomponent alloy of interest in light of crystal structures, elemental mixing, and phase separation. Here, we demonstrate the effectiveness of the proposed method by introducing physically meaningful phenomenological parameters that can be conveniently accessed from binary phase diagrams. These parameters are used to demarcate the phases forming regions for HEAs. The phases studied here are those with homogeneity ranges in the phase diagrams such as body-centered cubic (BCC) single-phase, face-centered cubic (FCC) single-phase, mixed FCC+BCC phase, hexagonal close-packed (HCP) single-phase, Sigma phase, and Laves phase. Minor phases such as line compounds are not included but will be for future work. A machine-learning (ML) algorithm is employed to navigate the complex parameter space regions occupied by the currently known HEA compositions. The effectiveness of the method is evaluated, and the derived ML algorithms are used to make predictions for experimental verification. The presented "phase diagram" approach to single solid solution HEAs can also complement CALPHAD and other first-principles methodologies in providing an efficient pathway to phase-field and microstructural control.

**Database Partitioning**

The HEAs included in our model have phases classified as: disordered FCC (A1), disordered BCC (A2), disordered HCP (A3), mixed disordered FCC+BCC (A1+A2), ordered BCC (B2), B2 mixed with disordered solid solution phases specifically A1, A2, and A3 (B2+SS), and either Sigma or Laves IM mixed with the other phases (IM+). The set of HEAs included in A1+A2 are the commingling of A1s, A2s, or the coexistence of A1s and A2s. The set of HEAs included in the IM+ phase have at least Sigma or Laves phase. Additionally, the IM+ phase may also contain other complex or solid solution phases. The database is parsed into three different levels, namely, Levels 1, 2, and 3. Level 1 is composed of the simple disordered phases: A1, A2, A1+A2, and A3. Level 2 is Level 1 with the addition of the B2+SS HEAs. And Level 3 is Level 2 with the addition of IM+ HEAs. HEAs with other minor phases such as line compounds that do not belong to the above categories are not included in the present study. Levels 1, 2, and 3 comprise 288, 416, and 529 HEAs respectively. More details about the database can be found in the method section and the supplementary materials.

**HEA Phase Formation Parameters**

The parameters, introduced below, and elaborated on in the method section, provide the basis for quantifying HEA phase formation tendencies. For ML, these individually measured property parameters used as input data to do classification are called features.

The HEA melting temperature ($T_m$) is expressed as the weighted average of binary liquidus temperatures. For the as-cast HEAs, undercooling usually extends to the region around 0.8 $T_m$[28]. Phase evolution may still exist below this temperature because of the high kinetic energies of the atoms. Here, a phase formation temperature ($T_{pf}$) is defined where rapid phase evolution ceases. It is assumed that $T_{pf}$ is not lower than 0.7 $T_m$. Below this temperature, the kinetic energy of atoms is not high enough to transform the phase within the brief time of cooling. Incidentally, most post-annealed HEAs in the full database are homogenized above 0.7 $T_m$. Atoms are free to exchange neighbors during undercooling (i.e. above 0.8 $T_m$), or via fast diffusion down to $T_{pf}$. The alloy mixture is essentially ergodic and local atoms have nearly equal probabilities to sample any binary configurations favored by the phases present in the constituent binary alloy diagrams.

Following the above discussion, information from individual binary phase diagrams is combinatorially used within the model. It is assumed that the probability for a pair of elements to form a specific phase is directly determined by its binary phase field percentage. The binary phase field percentage of phase X for i-j elemental pairs is denoted as $X_{i-j}$ and is determined using $T_{pf}$. $X_{i-j}$ is used to calculate the phase field parameter ($PFP_X$) which is the probability of a HEA to form a phase X.

Many mixed phase HEAs are found to form because of interatomic repulsions[29, 30]. Certain element pairs, such as Cr and Cu, separate because of the large positive mixing enthalpy, causing multiphase formations in HEAs[29]. This effect is included in the model with the phase separation parameter (PSP). Further details for $T_{pf}$ determination and calculating these parameters (value ≤ 1) are found in the method section.

**Visualization of Phase Regions in Parameter Space**

The prior defined parameters are calculated for all HEAs in different database levels. Their correlations with the actual phases formed are examined.

For the Level 1 phases, there are correlations between the calculated parameters $PFP_{A1}$, $PFP_{A2}$, $PFP_{A3}$, and PSP with the A1, A2, A3, and A1+A2 phase formation. Fig. 1a, a plot of $PFP_{A1}$ verse $PFP_{A2}$ shows the parameters partitioning the A1 and A2 HEAs. Typically, A1 HEAs have

$PFP_{A1} > 0.4$ and $PFP_{A2} < 0.4$, while A2 HEAs have $PFP_{A1} < 0.4$. Adding PSP as a third axis results in Fig. 1b, which separates out the A1+A2 HEAs from the A1 and A2 HEAs. A1+A2 HEAs are distributed in a region where neither $PFP_{A1}$ nor $PFP_{A2}$ is dominant. The relative higher PSP values differentiate them from the single A1 or A2 HEAs. In general, large $PFP_{A1}$ or $PFP_{A2}$ promotes the formation of a single phase, while the comparable $PFP_{A1}$ and $PFP_{A2}$ values tend to favor mixed phase formation. A large PSP causing phase separation will also lead to the A1+A2 phase formation. To study the effect of $PFP_{A3}$ on A3 phase formation, a plot with axes $PFP_{A1}$, $PFP_{A2}$, and $PFP_{A3}$ is plotted for A1, A2, A3, and A1+A2 HEAs in Fig. 1c, where A1, A2, and A1+A2 HEAs are grouped as non-A3 HEAs. All the A3 HEAs have higher $PFP_{A3}$ than the non-A3 HEAs and appear separate from the other phases.

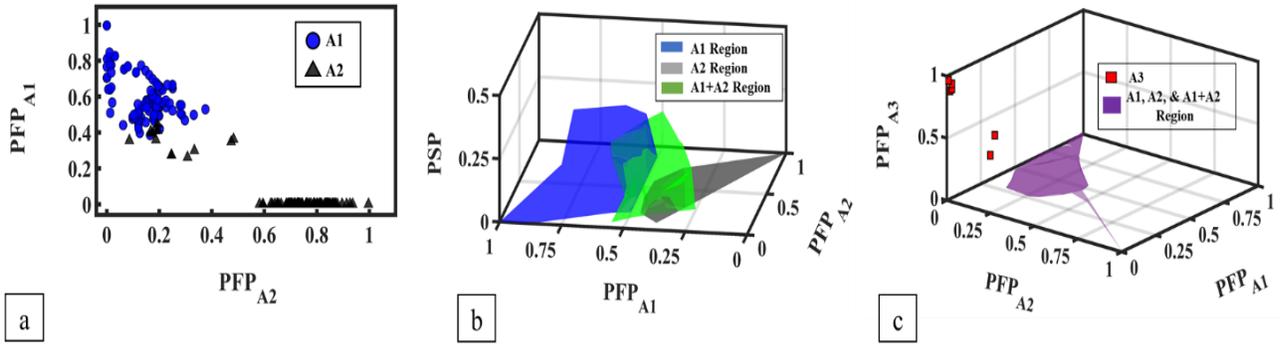

Fig. 1 Visualizations of Level 1 HEA parameters $PFP_{A1}$, $PFP_{A2}$, $PFP_{A3}$, and PSP for phases A1, A2, and A1+A2. (a) $PFP_{A1}$ is plotted against $PFP_{A2}$ for A1 and A2 HEAs; (b) $PFP_{A1}$, $PFP_{A2}$, and PSP are plotted for phase regions of A1, A2, and A1+A2 HEAs; and (c) $PFP_{A1}$, $PFP_{A2}$, and $PFP_{A3}$ are plotted for A3 HEAs and phase region of non-A3 (A1, A2, and A1+A2) HEAs.

For the Level 2 phases, the five parameters are $PFP_{A1}$, $PFP_{A2}$, $PFP_{A3}$, $PFP_{B2}$, and PSP. In Fig. 2a-g, to study the effects of $PFP_{B2}$, the 5D parameter space of the Level 2 data is visualized by projecting it on to 3D spaces. Fig. 2a is plotted with only the parameters in Level 1. B2+SS HEAs are mixed with HEAs in other phases. In Fig. 2b-d, $PFP_{B2}$ is added. Fig. 2e-g have the same axes as Fig. 2d but can give direct comparisons between the B2+SS phase and the A1, A2, and A1+A2 phases. On all these plots, B2+SS HEAs are located in a region with relatively higher $PFP_{B2}$ values. This indicates $PFP_{B2}$ is strongly correlated with the B2+SS phase formation. $PFP_{A3}$ and A3 HEAs are not plotted here because $PFP_{A3}$ has no effect on the formation of B2+SS phase and A3 HEAs are trivial to predict with $PFP_{A3}$ as shown in Level 1.

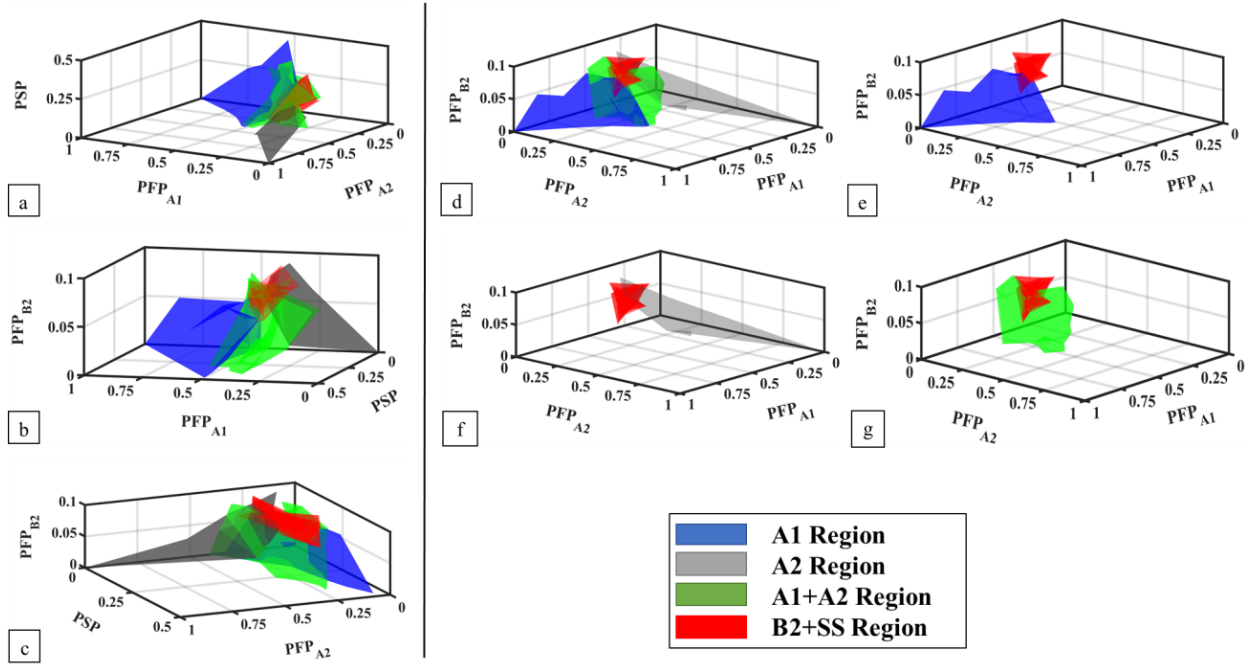

Fig. 2 Visualization of Level 2 parameters $PFP_{A1}$, $PFP_{A2}$, $PFP_{A3}$, $PFP_{B2}$, and PSP for the A1, A2, A1+A2, and B2+SS HEA phase regions. (a) $PFP_{A1}$, $PFP_{A2}$, and PSP; (b) $PFP_{A1}$, $PFP_{B2}$, and PSP; (c) $PFP_{A2}$, $PFP_{B2}$, and PSP; (d) $PFP_{A1}$, $PFP_{A2}$, and $PFP_{B2}$; and (e)-(g) the decomposition of the plot (d) highlighting the location of the B2+SS phase region relative to the A1, A2, and A1+A2 phase regions.

For the Level 3 phases, two additional parameters $PFP_{Sigma}$ and $PFP_{Laves}$ are added. 7 parameters $PFP_{A1}$, $PFP_{A2}$, $PFP_{A3}$, $PFP_{B2}$, $PFP_{Sigma}$, $PFP_{Laves}$, and PSP are used to separate the phase regions of A1, A2, A3, A1+A2, B2+SS, and IM+ HEAs. The newly added parameters $PFP_{Sigma}$ and $PFP_{Laves}$ are used to predict the appearance of Sigma and Laves IM phases. In order to study the correlation between the newly added IM+ phase formation and the two parameters $PFP_{Sigma}$ and $PFP_{Laves}$, a 2D graph with axes $PFP_{Sigma}$ and $PFP_{Laves}$ is plotted in Fig. 3. All the phases from Level 2 are grouped together as Non-IM phases. In general, IM+ HEAs have larger $PFP_{Laves}$ or $PFP_{Sigma}$ than most of the Non-IM HEAs. However, all 7 parameters have an influence on the IM+ phase formation. Fig. 3 is insufficient to convey all the information from the 7 parameters.

Level 1 shows separation between all single phase HEAs in the $PFP_{A1}$, $PFP_{A2}$, $PFP_{A3}$, and PSP parameter space. A1+A2 phase region is seen to have some overlaps with A1 and A2 phase regions. By adding more parameters in Level 2 and Level 3, additional overlaps are noted. The parameter space of the HEAs assumes an increasingly complex topological configuration as the number of parameters increases and it is difficult to resolve the connections in 3D space. In such complex cases, ML is superior to the visualization method to determine phase formation regions.

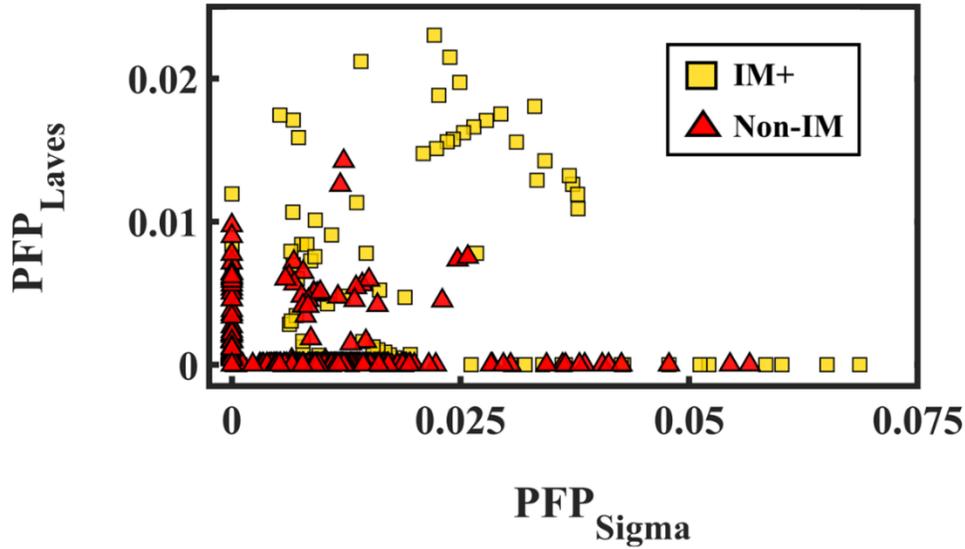

**Fig. 3** Level 3 parameters PFP$_{Sigma}$ and PFP$_{Laves}$ plotted for IM+ and Non-IM HEAs, where Non-IM includes A1, A2, A1+A2, and B2+SS.

**HEA Phases Prediction Using Machine Learning**

ML is employed to analyze the complex parameter space of HEA phase formation. It creates links in the multi-dimensional phase space that are not possible with the visualization method. Through ML composition-phase correlations are determined and new HEA compositions are predicted. As discussed in the method section, ML was conducted with features computed at various possible T$_{pf}$ ($\geq 0.7$ T$_m$) and the optimized T$_{pf}$ = 0.8 T$_m$ is obtained.

The effect of phase formation from alloy preparation methods is also studied. ML is first applied to only the as-cast HEAs and its performance serves as a benchmark. Then ML is applied to all HEAs in as-cast and annealed states. The ML prediction performance comparison of the two HEA sets yields on average that the addition of the annealed HEAs has a slight abating effect, as seen in Table 1.

The ML results for Level 1 HEAs are obtained using the features PFP$_{A1}$, PFP$_{A2}$, PFP$_{A3}$, and PSP. The overall success rates with 50 to 90 % training sets are 86-88 % for the as-cast HEAs or 85-86 % including the annealed HEAs. Single phase predictions have higher success rates than the mixed phase predictions. These high prediction success rates prove that these parameters are sufficient for describing the disordered solid solution phase formation behavior. PFP$_{B2}$ is added as a fifth ML feature to predict the B2+SS HEAs in Level 2. The overall and the B2+SS phase prediction success rates are near 80 % for both the as-cast HEA set and the set including the annealed HEAs. Thus, PFP$_{B2}$ is shown to be useful in predicting the presence of the B2+SS phase. Formation of the IM+ phases in the Level 3 HEAs are studied by adding PFP$_{Sigma}$ and PFP$_{Laves}$ as new features. The IM+ phase prediction success rates are 69-77 % for the as-cast HEAs or 68-74 % including the annealed HEAs. The overall success rate is as high as 75 % for all HEAs.

With the increasing complexity of the database from Level 1 to Level 3, the ML prediction success rates for all phase categories decrease. However, the prediction for single phase HEAs maintains an average success rate of about 80 % even at Level 3. B2 phase and IM phase formations are of certain interest in HEA design. The model has prediction success rates of about 75 % for these phases in Level 3.

Moreover, as the training set percentage changes from 90 % to 50 % at each level, the success rates show little variance. High accuracy is obtained even with training set percentage as low as 50 %.

| | | ML Prediction Success Rate (%) | | | | | | |
|---|---|---|---|---|---|---|---|---|
| | | As-Cast [As-Cast + Annealed] | | | | | | |
| Level | Phase Category / Training Set Percentage (%) | Overall | A1 | A2 | A1-A2 | A3 | B2+SS | IM+ |
| Level 1 | 90 | 88 [86] | 88 [86] | 93 [91] | 83 [81] | 100 [100] | | |
| | 80 | 87 [86] | 86 [86] | 91 [91] | 84 [81] | 100 [100] | | |
| | 75 | 86 [85] | 85 [84] | 92 [91] | 82 [81] | 100 [100] | | |
| | 67 | 86 [85] | 86 [85] | 92 [91] | 82 [80] | 100 [100] | | |
| | 50 | 86 [85] | 86 [86] | 91 [90] | 82 [80] | 100 [100] | | |
| Level 2 | 90 | 80 [80] | 87 [88] | 87 [89] | 68 [64] | 100 [97] | 80 [80] | |
| | 80 | 80 [80] | 85 [87] | 87 [88] | 69 [64] | 100 [100] | 79 [79] | |
| | 75 | 80 [79] | 87 [87] | 87 [87] | 67 [62] | 100 [100] | 81 [77] | |
| | 67 | 79 [79] | 85 [87] | 85 [88] | 69 [62] | 100 [93] | 79 [78] | |
| | 50 | 79 [80] | 86 [88] | 86 [88] | 65 [63] | 100 [95] | 78 [79] | |
| Level 3 | 90 | 74 [74] | 83 [82] | 75 [75] | 61 [59] | 100 [100] | 76 [77] | 76 [74] |
| | 80 | 75 [73] | 84 [82] | 75 [74] | 61 [59] | 100 [100] | 75 [76] | 77 [72] |
| | 75 | 73 [72] | 82 [80] | 74 [74] | 59 [58] | 100 [100] | 73 [75] | 74 [72] |
| | 67 | 72 [71] | 81 [80] | 75 [73] | 56 [57] | 100 [100] | 76 [75] | 70 [68] |
| | 50 | 70 [71] | 78 [80] | 75 [71] | 59 [56] | 100 [95] | 70 [74] | 69 [70] |
| | | Total Alloy Counts | | | | | | |
| | | As-Cast [As-Cast + Annealed] | | | | | | |
| Level | Phase Category | Overall | A1 | A2 | A1-A2 | A3 | B2+SS | IM+ |
| | Level 1 | 230 [288] | 79 [110] | 64 [79] | 83 [93] | 4 [6] | | |
| | Level 2 | 340 [416] | 79 [110] | 64 [79] | 83 [93] | 4 [6] | 110 [128] | |
| | Level 3 | 430 [529] | 79 [110] | 64 [79] | 83 [93] | 4 [6] | 110 [128] | 90 [113] |

Table 1: ML results and count of HEAs for the three levels of the study. ML prediction success rates for the as-cast HEAs and the as-cast + annealed HEAs in different phases are listed. The success rates are F1 scores. Counts of HEAs and phases for the as-cast HEAs and the as-cast + annealed HEAs in different phases are listed.

**Model Validation**

To show that the model avoids overfitting with ML and can expand the current phase regions, as shown in Fig. 4, 30 new HEAs were synthesized. The phases of these elemental combinations, not exist in the current collected database, are then predicted by the model. The selection of compositions is distributed evenly in parameter space of the collected database, which makes the numbers of new HEAs in different predicted phases approximately proportional to the numbers of different HEA phases in the database. The majority of our synthesized HEAs are outside the current known phase regions. As shown in Table 2, our method is not limited by the use of a specific element type nor the number of elements in a HEA. Elements are chosen from different groups of the periodic table such as refractory metals, transition metals, and main group elements. The number of elements in a single HEA varies from four to seven. All the phases are measured in as-cast states. Out of the 30 HEAs, 23 were predicted by ML correctly, yielding a success rate of 77 %. Their X-Ray Diffraction (XRD) patterns are found in the supplementary materials.

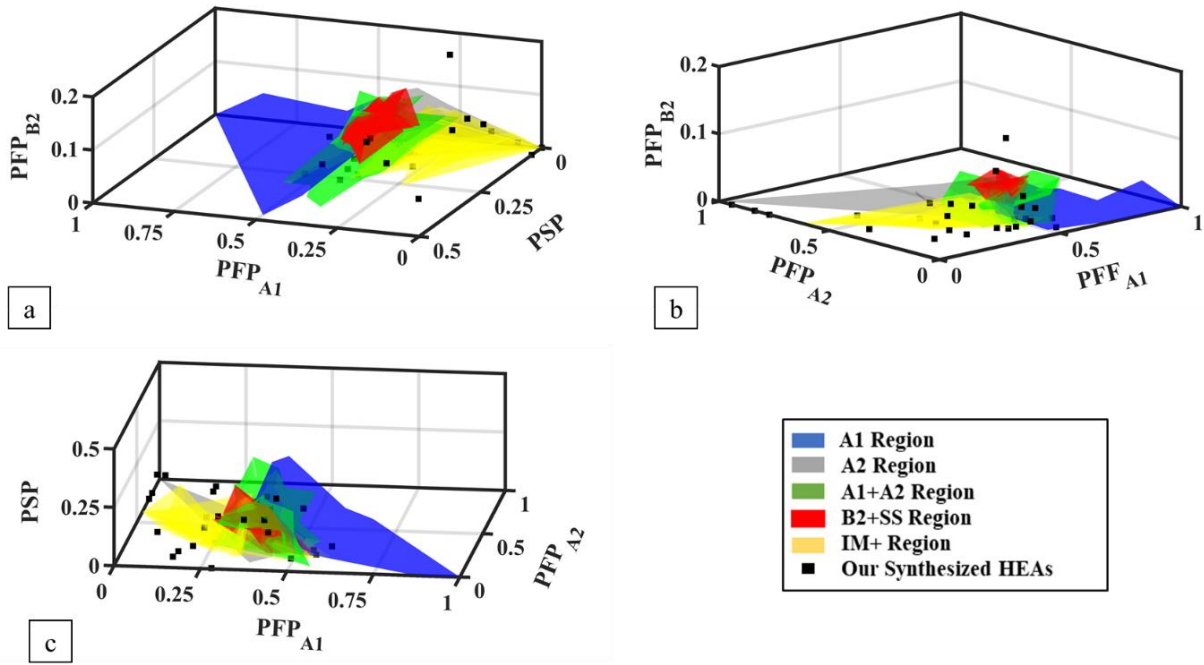

Fig. 4: Our synthesized HEAs locations are plotted relative to the phase regions of Level 3 for (a) $PFP_{A1}$, $PFP_{B2}$, and PSP; (b) $PFP_{A1}$, $PFP_{A2}$ and $PFP_{B2}$; and (c) $PFP_{A1}$, $PFP_{A2}$, and PSP parameters.

| Alloy (at.%) | Predicted Phase | Real Phase | Alloy (at.%) | Predicted Phase | Real Phase |
|---|---|---|---|---|---|
| $Ag_{0.2}Al_2CrMnNi$ | A1+A2 | **B2+A1+A2** | CoCrCuFe | A1+A2 | A1+A2 |
| AgAlCrMnNi | B2+SS | B2+A1+A2 | $CoCrCuMn_{0.8}Ti$ | IM+ | Laves+A1 |
| $Al_{0.5}CoCr_{0.5}CuMnNi$ | A1+A2 | **A1** | $CoCrFeMnNi_2V_{0.5}$ | A1 | A1 |
| $Al_{0.5}CoCuFeNiV_{0.5}$ | A1 | A1 | $CoCrFeMoNiV_{0.5}$ | IM+ | Sigma+A1 |
| $Al_2CoNb_{0.2}Ni$ | B2+SS | **Laves+B2+A2** | CoCrFeMoV | IM+ | Sigma |
| $AlCo_{0.5}CrCu_{0.2}FeMn$ | B2+SS | **A2** | $CoCrFeNb_{0.5}Ti_{0.5}$ | IM+ | Laves |
| $AlCo_2CrCuNi_3V$ | A1 | A1 | $CoCrFeNiSi_{0.6}$ | A1 | A1 |
| $AlCoCu_{0.5}Fe$ | B2+SS | B2+A2 | $CoCuFeMnNiV_{0.5}$ | A1 | A1 |
| $AlCoCuNiTi_{0.25}$ | B2+SS | B2+A1+A2 | $CoFeMnNiTi_{0.5}V_{0.5}$ | IM+ | **A1** |
| $AlCrCuFeNiSi_{0.25}$ | B2+SS | B2+A2 | $Cr_2FeNiTi$ | IM+ | Laves+A2 |
| $AlCrMoNi_3W_{0.5}$ | B2+SS | **A1+A2** | $CrCuFeMnNiTi_{0.3}$ | A1+A2 | A1+A2 |
| $Co_{0.2}TaTiV$ | A2 | A2 | CrNbNiTiZr | IM+ | Laves |
| $CoCr_{0.3}Cu_{0.2}FeNiV_{0.5}$ | A1 | A1 | CuFeMnNiV | A1+A2 | **Sigma+A1** |
| $CoCr_{1.5}Fe_{1.5}NiSi_{0.2}$ | A1 | A1 | $Hf_{0.5}NbTaW_{0.5}Zr$ | A2 | A2 |
| $CoCrCu_{0.5}FeNi_2Ti_{0.5}V_{0.5}$ | A1 | A1 | HfNbTaZr | A2 | A2 |

Table 2: HEAs synthesized to validate the ML model. The compositions, predicted phases by the ML in Level 3, and the XRD measured phases are listed. Recall that IM+ phase is the appearance of Sigma or Laves phase together with the potential existence of other phases such as the A1 and A2 solid solution phases. In the real phase column, the detailed phase information is listed. The seven HEAs whose measured phases differ from predictions are underlined. (XRD patterns cannot reveal if the B2 phase exists with or without the A2 phase because of diffraction peaks overlapping. Moreover, in HEAs B2 usually has a strong tendency to form with A2. Thus, B2 is listed together with A2 in the real phase information.)

**Discussion**

For the first time, a method predicting the phase formation of HEAs based solely on the binary phase diagrams is demonstrated and validated. The information on elemental mixing and phase separation from binary phase diagrams has provided success to the phenomenological approach presented. Considering the atomic mobility at high temperatures and presumed pairwise additivity of atomic pair interactions, this information from binary diagrams is used combinatorially to evaluate HEA phases formation. The initial success of using $PFP_X$ and PSP, defined using binary phase diagrams, in predicting the corresponding single phase and mixed phase HEAs, prompted us to apply this method to include more phases. The inter-correlated roles of these parameters are noted, and their combined effect must be considered in designing HEAs. We have included in our study the majority of the entire available HEA database, excluding 80 that contain line compounds and the minor phases. Visualization reveals robust HEA phase formation regions in the parameter space. ML enables the quantification of HEA phase formation, yielding an average prediction success rate > 85 % for the Level 1 and Level 2, and near 80 % for Level 3 for the single phases. The ML success rates obtained from the as-cast HEAs, or the as-cast and annealed HEAs vary marginally. Thus, the model works well for the as-cast and the high temperature annealed HEAs. Considering that these are the most common HEA preparation methods, our model can be applied to most HEA synthesis situations. High accuracy is obtained even with small training set percentages. This implies that the phase formation parameters are well defined and efficient in prediction. The resulting success rates are validated experimentally. Moreover, ML can predict the phases of the new HEAs to expand the current database and phase parameter regions.

Compared with the other large database statistical approaches, Tancret et al. combined Gaussian Process using nine thermodynamic and atomistic parameters with CALPHAD to predict the formation of over 60 single solid solution phase HEAs.[26] The performance of the model has high precision but low recall. Many of the alloys predicted as single solid solution HEAs by this method have a high chance of being single solid solution HEAs, but many potential single solid solution HEAs are misidentified as mixed phases HEAs. Additionally, the exact phase of a HEA such as BCC or IM cannot be predicted. As a comparison, our method has high precision and high recall, and gives specific phase formation information.

Another model by Kube et al. assigned values called stabilizing abilities ($\beta_i$) to seven specific elements Al, Co, Cr, Cu, Fe, Mn, and Ni representing their strength in stabilizing FCC or BCC formation. The $\beta_i$'s are optimized by ordinal logistic regression based on a database of over 2000 sputter deposited HEAs from a high-throughput experiment.[27] This method is efficient in separating out FCC and BCC single phase HEAs. But mixed FCC and BCC phase cannot be separated from the prior phases. Moreover, other phases such as HCP and IM were not studied. Our method has no element preference and more phases can be predicted.

To summarize, the advantages of our approach are the following:

1. Indiscriminate HEA selection feasibility: Some prediction methods such as CALPHAD are limited by the availability and depth of proprietary databases. Our method is based solely on binary phase diagrams for which there exist plentiful easily accessible data.

2. Phase region expansion ability: New HEAs are predicted with a high success rate outside the regions where more than 500 HEA phases are currently known.

3. Ease of computing: Methods such as ab initio molecular dynamics require high computation capability. This model can be run on a laptop, no high-performance computing facilities are needed.

**Acknowledgment:**
The authors appreciate the assistance from Adam Summers while at the University of Virginia in building the database. This work is supported by Office of Naval Research grant N00014-18-1-2621.

## Method

Melting Temperature:

The $T_m$ is calculated from the liquidus temperatures in binary phase diagrams. $c_i$ and $c_j$ are the atomic percentages of the elements i and j. For the binary pair i-j, binary liquidus temperatures $T_{i-j}$ can be found at the composition where i and j element relative ratio is $c_i : c_j$. $T_m$ of the whole system will be calculated by the following equation (1):

$$T_m = \frac{\sum_{i \neq j} T_{i-j} \times c_i \times c_j}{\sum_{i \neq j} c_i \times c_j} \qquad (1)$$

where the summation is over all the i-j pairs in the alloy system.

Methods of Calculating Parameters:

*Calculating $PFP_X$:* The method of calculating binary phase field percentage, $X_{i-j}$, uses line segments at $T_{pf}$. $X_{i-j}$ is the percentage of the line segment between the two intersection points of an isotherm at $T_{pf}$ and the phase boundary of phase X. It is assumed that the phases at solidification are directly related to the phases occurring at $T_{pf}$ because the phase transformation occurs for a longer duration near $T_{pf}$ as opposed to near $T_m$ due to a decreasing cooling rate when the temperature is decreased.

$PFP_X$ is calculated from $X_{i-j}$ by equation (2), where $c_i$ and $c_j$ are the atomic percentages of i-th and j-th elements.

$$PFP_X = \frac{\sum_{i \neq j} X_{i-j} \times c_i \times c_j}{\sum_{i \neq j} c_i \times c_j} \div 100 \% \qquad (2)$$

An example of $PFP_X$ calculation is shown in Fig. M1. Here the Cr-Ni phase diagram is used to calculate $A2_{Cr-Ni}$ and $A1_{Cr-Ni}$. Using the HEA $Al_2CoCrCuNi$, with a $T_m = 1569$ K, the phases are assumed to form at $T_{pf} = 1255$ K and the method gives $A2_{Cr-Ni} = 5\%$ and $A1_{Cr-Ni} = 44\%$.

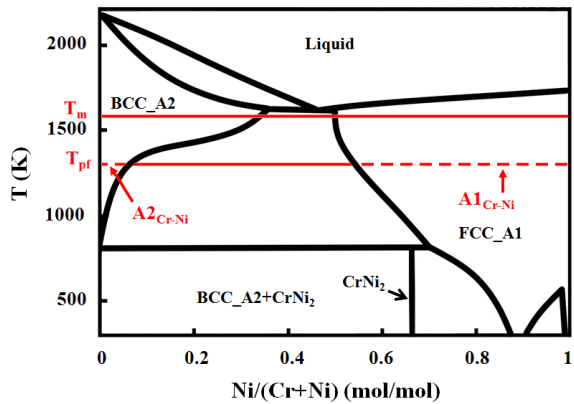

Fig. M1: Demonstration of the binary phase field percentage calculation. The binary phase diagram Cr-Ni is used to determine the $A1_{Cr-Ni}$ and $A2_{Cr-Ni}$ for the HEA $Al_2CoCrCuNi$.

*Calculating PSP:* The binary phase separation percentage for atomic pairs i and j, $Separation_{i-j}$, is calculated using the line segment method at $T_{pf}$. The remaining phase percentage is $Mixing_{i-j}$. The PSP for a HEA is defined by equation (3):

$$PSP = \frac{\sum_{i \neq j} Separation_{i-j} \times c_i \times c_j}{\sum_{i \neq j} Mixing_{i-j} \times c_i \times c_j} \quad (3)$$

with $Mixing_{i-j} = 1 - Separation_{i-j}$.

The atomic pairs with the separation effect are identified on phase diagrams by the presence of two bounding pure solid solution phases with no additional single phases present between the two. For example, a strong phase separation effect exists on the phase diagram of Cr-Cu (Fig. M2a) where Cr and Cu never dissolve into the same phase matrix.

In certain cases, at high temperatures, the negative mixing entropy term is large enough to overcome the positive mixing enthalpy and results in a negative Gibbs free energy for forming the solid solution. This makes it possible to have the two elements mixed marginally. Co-Cu in Fig. M2b is a typical example where two atoms separate at low temperature and mixing exists at high temperature. The Co-Cu phase diagram is used to calculate the $Separation_{Co-Cu}$ by the line segment method. The HEA $Al_2CoCrCuNi$ is used again. This method gives a $Separation_{Co-Cu}$ = 92 % and $Mixing_{Co-Cu}$ = 8 %. $Separation_{i-j}$ = 0 % if the phase separation is absent from a phase diagram.

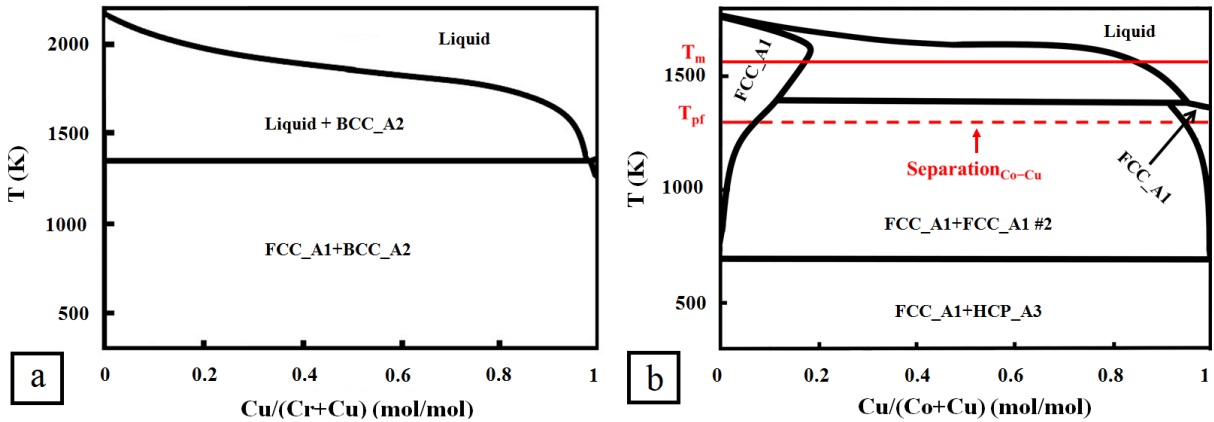

Fig. M2: Two binary phase diagrams used to determine binary phase separation percentage for HEA $Al_2CoCrCuNi$. (a) Phase diagram of Cr-Cu to show a complete phase separation effect. (b) Overlay of the Co-Cu phase diagram illustrating the line segment method to determine the $Separation_{Co-Cu}$ for the HEA $Al_2CoCrCuNi$.

For an as-cast HEA, the phase transformation evolves at various temperatures above $T_{pf}$ as it cools from the molten state. $PFP_X$ and PSP are calculated from values determined by the line segment method using various $T_{pf}$ values from 0.7 to 0.9 $T_m$. $T_{pf}$ is then determined by optimizing the ML results. The optimal results are obtained when $T_{pf} = 0.8\ T_m$. Of note, the optimized $T_{pf}$ is close to the undercooling temperature.

For the high temperature annealed HEAs, the phases formed during annealing at these high temperatures are locked in during rapid quenching. Thus, $T_{pf}$ is the annealing temperature and the phase formation tendency is determined from the line segment percentages of the binary phase fields present.

Machine Learning:

ML was conducted using the data mining software WEKA 3.8[31]. We use Random Forest[32] with 300 trees to perform this classification task. The features are the parameters defined for the three levels of the database partition. Each database level is divided randomly into training and test sets. The ML algorithm establishes and optimizes decision trees based on the training set. These trees are used to predict the phases of HEAs in the test set based on their features. The performance of the ML model is accessed by 2, 3, 4, 5, and 10-fold cross-validations, which, in Table 1, correspond to training set percentages of 50 %, 67 %, 75 %, 80 %, and 90 %. An F1 score, as a weighted average of precision and recall model evaluation metrics, is used to denote the success rate of prediction. Each cross-validation is conducted for 20 times and then the average F1 score is obtained. After the optimization, new HEAs are predicted.

Alloy Validation Experiment:

The 30 predicted HEAs used to validate our model were all prepared by suction casting. These HEAs are created by first making master ingots. These ingots are made from elements with a minimum purity of 99.7 wt%. The elements are arc-melted in a water-cooled copper hearth in a high purity argon atmosphere and are melted three times to ensure homogeneous mixing. The ingots are then suction-casted into a copper mold making 3 mm diameter rods. Structure investigations are carried out with XRD analysis using a Cu Kα radiation on a PANalytical Empyrean diffractometer.

Database Description:

More than 600 HEAs have been collected from literature[6, 33-135] in supplementary notes. Structural data used in our model is predominantly from XRD measurements. When transmission electron microscopy (TEM) data is available and it can reveal the hidden patterns from XRD results, the higher resolution TEM data will supersede the XRD data. The 529 HEAs studied are focused on those formed in as-cast state or those annealed at temperatures that are higher than 0.7 $T_m$, as most of the heat-treated HEAs were annealed above it. Also, the reason for including high-temperature annealing HEAs is that the formation entropy can contribute more Gibbs free energy change when the HEA is annealed at a higher temperature. Mechanically alloyed HEAs are not included because ball milling has the tendency to retain metastable phases.